Similarities and differences in the sensitivity of Soil Organic Matter (SOM) dynamics to biogeochemical parameters for different vegetation inputs and climates

Authors: Ceriotti, G.[a,b]; Tang, F. H. M.[b] & Maggi, F.[b]

[a] Institute of Earth Sciences, Building Geopolis, University of Lausanne, Lausanne, Switzerland.
[b] Laboratory for Environmental Engineering, School of Civil Engineering, The University of Sydney, Bld. J05, 2006, Sydney, NSW, Australia.

Corresponding Author: Giulia Ceriotti, giulia.ceriotti@unil.ch

**Abstract**

The biogeochemical complexity of environmental models is increasing continuously and model reliability must be reanalysed when new implementations are brought about. This work aims to identify influential biogeochemical parameters that control the Soil Organic Matter (SOM) dynamics and greenhouse gas emissions in different ecosystems and climates predicted by a physically-based mechanistic model. This explicitly accounts for four pools of organic polymers, seven pools of organic monomers, five microbial functional groups, and inorganic N and C species. We first benchmarked our model against vertical SOM profiles measured in a temperate forest in North-Eastern Bavaria, Germany (Staudt and Foken, 2007). Next, we conducted a sensitivity analysis to biogeochemical parameters using modified Morris indices for target SOM pools and gas emissions from a tropical, a temperate, and a semi-arid grassland in Australia. We found that greenhouse gas emissions, the SOM stock, and the fungi-to-bacteria ratio in the top soil were more sensitive to the mortality of aerobic bacteria than other biogeochemical parameters. The larger $CO_2$ emission rates in forests than in grasslands were explained by a greater dissolved SOM content. Finally, we found that the soil N availability was largely controlled by vegetation inputs in forests and by atmospheric fixation in grasslands.

**List of Abbreviation**

AER = Aerobic Bacteria

AmA = Amino-Acids

AmS = Amino-Sugar

AOB = Ammonia Oxidizing Bacteria

BAMS2 = Biotic and Abiotic Model of SOM version 2

C = Carbon

Cls = Celulose

DEN = Denitrifying Bacteria

F = Fungi

GSA = Global Sensitivity Analysis

HCls = Hemi-Celulose

Lig = Lignin

Lip = Lipids

LSA = Local Sensitivity Analysis

Msa = Monosaccarides

N = Nitrogen

NOB = Nitrite Oxidizing Bacteria

Nti = Nucleotid

OAT = One-factor At Time method

OraA = Organic Acid

Pgl = Peptidoglycan

Phe = Phenols

SAG = Semi-Arid Grassland

SOM = Soil Organic Matter

TEF = Temperate Forest

TEG = Temperate Grassland

TRG = Tropical Grassland

# 1 Introduction

The capability of environmental models to capture the complexity of nutrient biogeochemistry within ecosystems has deeply been improved over the past years; yet, their reliability must be re-assessed when applied to specific aspects of nutrients dynamics and ecosystem responses to interventions, to aid in interpreting the ecosystem transitions caused by shifts in climatic patterns and related feedbacks, or when new capabilities are implemented. Among the multiple dimensions that describe the biogeochemistry of nutrients, Carbon (C) and Nitrogen (N) dynamics are of particular interest for their repercussions on greenhouse gas exchanges with the atmosphere. Soil Organic Matter (SOM) consists of 2,157 to 2,293 Pg-C and 133 to 140 Pg-N in the upper 100 cm of soil (Batjes, 2014) and it plays a crucial role in ecosystem stability. Despite a wide research around SOM dynamics in the last decades, the factors controlling the soil ability to act as a source or as a sink of greenhouse gases is still uncertain due to the interdependence of underpinning biotic and abiotic processes (Armstrong et al., 2015); specifically, the soil C and N dynamics emerge from a complex network of biogeochemical processes including organic matter-mineral interaction, plant inputs, and microbial activity. Thus, reaching a deeper understanding and improving the predictions of soil nutrients dynamics by robust modeling is essential to provide information to policy makers for designing adaptation and mitigation strategies.

Many SOM models (e.g., Century, Parton et al. (1988), or RothC, Jenkinson and Coleman (2008)) organize the organic matter into a suite of pools that exchange C by means of first-order kinetic processes. The latter use effective parameters typically estimated through observations aiming to embed the effect of molecular recalcitrance, microbial activity, and soil protection. The lack of explicit mechanistic representation of physical processes limits the scope of those models to explore ecosystem responses in conditions that differ from those associated with parameter estimation (Lehmann and Kleber, 2015; Treseder et al., 2012). Alternatively, physically-based models embedding diverse levels of complexity have been developed to mechanistically describe SOM dynamics in a generalized form (Riley et al., 2014; Ahrens et al., 2015; Riley, 2013; Allison et al., 2010; Fontaine and Barot, 2005; Lawrence et al., 2009; Gerber et al., 2010; Moorhead and Sinsabaugh, 2006; Sierra et al., 2012; Achat et al., 2016) and are supported by increasing computational power. The comparative assessment by Lawrence et al. (2009) has indeed demonstrated the richer information provided by mechanistic as compared to traditional, low-complexity or simplistic models. Major shortcomings are the high computational cost and a large number of parameters often difficult to estimate (Lawrence et al., 2009; Sivakumar, 2008). For these reasons, including a complex SOM mechanistic model into an ecosystem model is generally considered unfeasible (Riley, 2013). It is therefore essential to develop approaches that can capture the salient features of SOM dynamics by finding the best compromise between computational costs and the need to include a large number of processes and parameters (Riley, 2013; Sivakumar, 2008).

Sensitivity analyses have proved useful to explore influential parameters in complex SOM models (e.g., Wang et al., 2013; Yu et al., 2013; Ahrens et al., 2015), provide information for model reduction, and drive model simplification. Different methods are currently available and their theoretical foundations and robustness in practical applications have been reviewed comprehensively in a number of works (e.g., Menberg et al., 2016; Dell'Oca et al., 2017; Pianosi et al., 2016; Rakovec et al., 2014; Razavi and Gupta, 2015; Saltelli et al., 2008; Campolongo et al., 2011; Cariboni et al., 2007; Tian, 2013). However, the majority of sensitivity analyses applied to SOM models have used deterministic frameworks (e.g., Larocque et al., 2008) or Local Sensitivity Analysis (LSA) methods (e.g.,One-factor At Time, OAT). These explore model outputs sensitivity (gradients) only in the neighborhood of selected parameter combinations. Global Sensitivity Analyses (GSA), instead, investigate the sensitivity of model outputs across an arbitrary range of the parameter space. GSA has already found useful implementations in hydrology (e.g., Ciriello et al., 2013; Song et al., 2015; Dell'Oca et al., 2017), geology (e.g., Formaggia et al., 2013; Colombo et al., 2017, 2018) and geochemistry (e.g., Ceriotti et al., 2018; Porta et al., 2018), and we contend that GSA can be extended to SOM biogeochemistry too. The higher informative content of GSA typically implies a high computational cost to run the model to exhaustively explore the parameter space. Hence, estimating GSA metrics is not computationally affordable for complex models embedding many parameters in this work. In these cases, screening methods such as the one proposed in Morris (1991) represent an effective compromise between the information provided by GSA and the efficient LSA procedures that avoid thousands of model runs (Menberg et al., 2016; Campolongo et al., 2011; Cariboni et al., 2007; Campolongo et al., 2007). The reliability of the Morris method has been tested and discussed in many works (Sumner et al., 2012; Campolongo et al., 2011, 2007; Cariboni et al., 2007; Herman et al., 2013; Iooss and Lemaître, 2015; Menberg et al., 2016; Pianosi et al., 2016), and we consider this approach appropriate to our purpose.

Here, we conduct a parameter screening of SOM modeling to identify key biogeochemical parameters controlling soil C and N dynamics in different climates and ecosystems. To this end, we have deployed the BAMS2 (Biotic and Abiotic Model of SOM version 2, Tang et al., 2019) mechanistic SOM model that explicitly accounts for biological, chemical, and physical processes regulating C and N mass exchanges between the soil, plant, and atmosphere. BAMS2 de-scribes the time evolution of four pools of organic polymers, seven pools of organic monomers, five microbial functional groups, inorganic C and N compounds as well as SOM protection, and plant nitrogen and water uptake. BAMS2, integrated in a general-purpose solver (BRTSim 3.1a, Maggi, 2019), was tested against field data from a temperate forest located in Germany and compared to a previously published mechanistic SOM model (i.e., COMMISSION, Ahrens et al., 2015) that differs from what we propose here. We next used the benchmarked modeling framework to analyse SOM dynamics in a tropical, a temperate, and a semi-arid grassland in Australia. We conducted a screening of the biogeochemical parameters by means of a modified Morris method (Morris, 1991) targeting soil $CO_2$, $NH_3$, $N_2O$ and NO gas exchanges with the atmosphere, and the total C and N stocks in the root zone as outputs of interest. This allowed us to identify influential biogeochemical parameters for the selected model outputs and critically discuss: (*i*) the parameters mostly contributing to model output variance and hence requiring careful experimental estimation; and (*ii*) SOM dynamics in different ecosystems including how biological processes affect the system dynamics as compared to others factor such as climate and vegetation inputs.

## 2  Methods and Materials

### 2.1  The coupled C-N model

We used the SOM mechanistic model BAMS2 in Tang et al. (2019), which couples the C cycle model BAMS1 by Riley et al. (2014) and the N cycle by Maggi et al. (2008). The SOM reaction network in BAMS2 includes: four non-soluble SOM polymeric pools, i.e., Lignin (Lig), Cellulose (Cls), Hemi-Cellulose (HCls) and Peptidoglycan (Pgl); seven soluble SOM monomeric pools, i.e., Monosaccharides (Msa), Amino-Acids (AmA), Amino-Sugar (AmS), Nucleotides (Nti), Phenols (Phe), Organic Acids (OrA) and Lipids (Lip); and seven inorganic N species ($NH_3$, $NH_4^+$, $NO_3^-$, $NO_2^-$, $NO_2$, NO, $N_2O$). The size of these SOM and N pools are regulated by chemical, physical, and biological processes outlined in Figure 1. All reactions in BAMS2 are detailed in the Online Resource (Section S.1, Table 4), while we illustrate the key model features below.

*Gaseous-aqueous mass exchange*
BAMS2 accounts for the dynamics of both gaseous and liquid phases in soil; the exchange of $O_2$, $CO_2$, $NH_3$, NO, $N_2O$, $N_2$ species between these phases is modeled according to a local thermodynamic equilibrium expressed as a function of temperature as prescribed in Maier and Kelley (1932) and Parkhurst and Appelo (2013).

*Microbial activity*

We accounted for heterotrophic aerobic fungi (F) and bacteria (AER), ammonia-oxidizing bacteria (AOB), nitrite-oxidizing bacteria (NOB) and denitrifying bacteria (DEN). These microbial groups were stoichiometrically defined as $C_{5x}H_{8x}O_{2x}N$ with $x = 1$ for bacteria and $x = 1.6$ for fungi according to the C:N ratios suggested by Mouginot et al. (2014).

Heterotrophic fungi (F) represent all decomposers with high enzymatic activity and low metabolic nutrient demand that depolymerize recalcitrant substrates (e.g., Fabian et al., 2017) such as Lignin (Lig), Hemi-Cellulose (HCls) and Cellulose (Cls) and release inorganic N and C, Monosaccharides (Msa) and Phenols (Phe). Heterotrophic bacteria (AER) depolymerize Pgl and mineralize all monomeric SOM compounds during growth and respiration. The N in SOM monomers is partially assimilated by AER during growth and partially released in the environment as free $NH_4^+$. AOB oxidize $NH_4^+$ to $NO_2^-$, which is used by NOB to produce $NO^-$; both AOB and NOB use $HCO_3^-$ as the C source in these reactions. Finally, DEN produce $N_2$ in four sequential reduction steps ($NO_3^- \rightarrow NO_2^- \rightarrow NO \rightarrow N_2O \rightarrow N_2$) in anoxic conditions.

Growth of all microbial functional groups is limited by space, water and substrate availability. All functional groups undergo mortality and the necromass decomposition returns polymeric (Pgl) and monomeric (Msa, AmA, AmS, OrA, Lip and Nti) compounds to the soil, which feed the C and N cycles anew. Microbial dynamics were modeled using Michaelis-Menten-Monod kinetics as

$$\frac{d[B]}{dt} = \sum_j Y_{j,B} R_{j,B} - \delta_B [B] \qquad (1)$$

where $[B]$ represents the concentration of a generic microbial group $B$ among F, AER, AOB, NOB, and DEN, $\delta_B$ is the mortality rate of $B$; $Y_{j,B}$ is the biomass yield of $B$ for the $j^{th}$ reaction contributing to $B$ growth, while $R_{j,B}$ is the rate of the $j^{th}$ reaction involving $B$, which is calculated as

$$R_{j,B} = k_j \cdot f_S \cdot \frac{[B]}{Y_{j,B}} \cdot \prod_i \frac{[X_i]}{[X_i] + K_{Mi}} \cdot \prod_m \frac{K_{Im}}{K_{Im} + [X_m]} \qquad (2)$$

where $k_j$ is maximum rate, $K_{M,i}$ is the Michaelis-Menten half-saturation constant of substrate $i$ and $[X_i]$ the concentration of the substrate $i$; $[X_m]$ is the concentration of the $m$ inhibitor with $K_{I,m}$ as inhibition constant, and $f_S$ quantifies the inhibition of soil moisture stress on the microbial activity (Online Resource, S2.1).

*Chemo-denitrification*

Abiotic denitrification can occur in acidic conditions and can compete with biological denitrification. The chemo-denitrification rate $R_{cd}$ is modeled as

$$R_{cd} = k_{cd} \frac{[NO_2^-]}{[NO_2^-] + K_{M,NO_2^-(cd)}} \cdot \frac{[H^+]}{[H^+] + K_{M,H^+}} \qquad (3)$$

where $k_{cd}$ is the maximum rate constant, while $K_{M,NO_2^-(cd)}$ and $K_{M,H^+}$ are the half-saturation constants of $NO_2^-$ and $H^+$, respectively.

*Nutrients input and plant uptake*

We used Michaelis-Menten kinetics to account for plant $NO_3^-$ and $NH_4^+$ uptake, and first-order kinetics for SOM inputs to soil through litter and roots exudate. Specifically, wood and leaf litter, composed of Cls, HCls, Msa, Lip, AmA, Nti (only for leaf litter) and Phe (only for wood litter), were assumed to have C:N = 35 (Moretto et al., 2001; Thomas and Asakawa, 1993) and introduce monomers and polymers from aboveground to the top soil. Root exudates, composed of Msa, Lip, OrA and AmA (Grayston et al., 1997), were assumed to have C:N = 12 (Mench and Martin, 1991; Grayston et al., 1997) and were distributed throughout the soil profile proportionally to a negative exponential distribution function describing the roots density profile (Ahrens et al., 2015).

$N_2$ fixation was described in BAMS2 by zero-order kinetics and was inhibited by soil moisture stress similarly to the microbial activity. $N_2$ fixation feeds the topsoil with dissolved $NH_4^+$ along the soil profile proportionally to the roots distribution at the maximum rate constant $k_{N-fix}$.

*Protection*

We identify protection as an ensemble of complex chemo-physical processes (Luo et al., 2017) that cause organic compounds to stabilize and age in soil (Riley et al., 2014; Six et al., 2002; Luo et al., 2017). It is generally recognized that SOM can be protected from enzymes and microbial activity through micro-aggregation and adsorption on silt and clay particles (Six et al., 2002). Evidence suggests that the protected SOM pool size is finite and depends on various soil physico-chemical characteristics. A broadly accepted physically-based model of SOM protection is still lacking (Luo et al., 2017), and we relied on Langmuir kinetics similar to the one used in Ahrens et al. (2015) and written as

$$\frac{d[X_{(p)}]}{dt} = k_{p,X} \cdot (Q_{max} - [X_{(p)}])[X_{(aq)}] - k_{unp,X} \cdot [X_{(p)}] \qquad (4)$$

where $Q_{max}$ is the maximum soil protection capacity, $k_{p,X}$ and $k_{unp,X}$ describe the protection and unprotection rates of the compound *X*, respectively, while [$X_{(p)}$] and [$X_{(aq)}$] are the concentrations [mol/L] of the protected and aqueous form of the generic chemical *X*, respectively.

## 2.2 BAMS2 implementation

The BAMS2 reaction network was implemented in the BRTSim v3.1a general-purpose solver described in Maggi (2019). BAMS2 included 56 reactions comprising 70 biogeochemical parameters. The values of those associated with the C cycle (R1 to R11, R39, R40, R44 to R46, R54 and R56 in Table 4, Online Resource) were taken from Riley et al. (2014), while those associated with the N cycle (R12 to R18, R41 to R43 and R47 to R53 in Table 4, Online Resource) were taken from Maggi et al. (2008). The kinetic protection rate $k_p = 3.30 \times 10^{-12}$ L mol s$^{-1}$ used here implied that protection competed with bacteria for substrate uptake, while the unprotection kinetic rates $k_{unp}$ were estimated from the $k_p/k_{unp}$ ratios proposed by Riley et al. (2014) for the different monomers and resulted in some compounds be preferentially protected as compared to others (R19 to R38 in Table 4, Online Resource). All parameters are listed in Online Resource (Tables 4 and 5) along with their reference values. Modeling details specific to BAMS2 integration into BRTSim such as grid, boundary conditions, and others are provided in Section 2.4.

## 2.3 Sensitivity analysis and target variables

The method proposed here for parameter screening makes use of the Elementary Effect (EE) Morris method (Campolongo et al., 2007), which is briefly recalled below. First, we assumed that each biogeochemical parameter $p_k$ can vary in a uniform interval $U_k$ spanning between ±50% of the benchmark reference parameter values (Table 5, Online Resource). For the aqueous-gaseous equilibrium constant, instead, we assumed that these parameters are described by polynomials of the temperature and not by a single parameter and the procedure implemented to account for their uncertainty is detailed in Section S.3, Online Resource. Each interval $U_k$ was normalized to [0,1] and then divided into $l = 11$ levels evenly spaced by $\Delta_k$. The parameter space corresponding to $K$ parameters used in BAMS2 is a $K$-dimensional hypercube $H^K$; any points within $H^K$ is identified by a vector of $K$ elements, each representing one parameter of our model. The parameter space $H^K$ was then explored by trajectories of points generated as follow:

1. the trajectory origin was selected by sampling one of the $l$ levels of each parameter $p_k$ randomly;

2. the trajectory was obtained by moving one parameter at a time by $\Delta_k$. The resulting trajectory was, therefore, a sequence of $K+1$ points in $H_K$;

3. The BAMS2 model was run to calculate the target output $y$ for as many points as within the trajectory;

4. The Elementary Effect $EE_{k,y}$ for parameter $p_k$ associated with $y$ was calculated as

$$EE_{k,y} = \frac{y(p_1, ..., p_k + \Delta, ...p_K) - y(p_1, ..., p_k, ...p_K)}{\Delta_k} \times \frac{1}{\sigma_y} \qquad (5)$$

where $\sigma_y$ is the standard deviation of $y$.

New trajectories are generated by replicating the steps above $n$ times and are used to calculate the corresponding $EE^i_{k,y}$ for each $i = 1, ..., n$ trajectory. The sensitivity indices $\mu_{k,y}$ are then calculated for each parameter $k$ and the set of $n$ values of $EE^i_{k,y}$ as

$$\mu_{k,y} = Median|EE_{k,y}| \qquad (6)$$

Note that Eqs. (5) and (6) differ from the original metrics proposed in Morris (1991), that is, $EE_{k,y}$ in Eq. (5) is normalized to $\sigma_y$ after recommendations in Sumner et al. (2012) to allow for a comparison of $\mu$ in Eq. (6) across different target outputs $y$. In addition, $\mu$ was calculated using the median of $EE_{k,y}$ as suggested by Menberg et al. (2016) instead of the mean to reduce the impact of outliers and enhance Eq. (6) robustness for small $n$ (Eq. (6) can be considered stable for $n \geq 10$). Because a single model realization can take 2 to 50 hours on a Intel(R) Core(TM) i7-6700 CPU @ 3.40GHz, we used $n = 10$, which required 710 model runs for the analysis of each case study investigated in this work. Finally, Eq. (6) uses the absolute value $|EE_{k,y}|$ to avoid changes in sign as prescribed in Campolongo et al. (2011).

The selected output variables for the parameter screening are: (*i*) the average soil daily emission rate of $CO_{2(g)}$ [kg-C/day/m$^2$], $NH_{3(g)}$, $N_2O_{(g)}$ and $NO_{(g)}$ [kg-N/day/m$^2$]; and (*ii*) the total organic carbon (C-stock) [kg-C/m$^2$], organic nitrogen (N-stock) [kg-N/m$^2$], and inorganic nitrogen ($N_{inorg}$-stock) [kg-N/m$^2$] in the top 30 cm of soil for comparison with Ahrens et al. (2015). The screening was conducted on four case studies described below with different vegetation inputs and climates.

## 2.4 Case studies and modeling settings

We used SOM observations from a temperate conifer forest (50°08'32" N and 11° 52'01" E) at the Coulissenhieb (I) site in Bavaria, Germany, described in Staudt and Foken (2007) to benchmark BAMS2. Specifically, observed SOM vertical profiles to 80 cm depth included: (*i*) total and protected organic C in [kg-C $m^{-3}_{soil}$] ($C_{(tot)}$ and $C_{(p)}$, respectively) reported in Ahrens et al. (2015); *(ii)* microbial biomass in [kg-C m$^{-3}_{soil}$] reported in Ahrens et al. (2015); and *(iii)* total organic N in [kg-N m$^{-3}_{soil}$] reported in Staudt and Foken (2007). In this case study, BRT-Sim was run for a soil column of 3 meters discretized into 20 nodes with free drainage bottom boundary condition. The stratigraphy and soil composition are reported in Table 1 (Staudt and Foken, 2007). The soil textural fractions were used to estimate the Brooks-Corey hydraulic parameters (Brooks and Corey, 1964) using the empirical scaling laws in Clapp and Hornberger (1978). The root vertical distribution was assumed to be negative exponential with 7.5 cm mean depth (Ahrens et al., 2015). Historical daily precipitation and minimum and maximum temperature from 2007 to 2017 were obtained from the DWD Climate Data Center, CDC (Source: Deutscher Wetterdienst, Kaspar et al., 2013). These data were used to generate synthetic daily precipitations and temperatures using a Richardson-type weather generator (Chen et al., 2010). The potential evapotranspiration $ET_0$ was calculated using the Food and Agriculture Organization (FAO) tool (Allen et al., 1998), while the actual evapotranspiration was calculated as $ET_c = k_c \cdot ET_0$, where $k_c = 1$ is the proportionality coefficient for a conifer forest suggested by the FAO (Allen et al., 1998). We verified that the generated precipitations were statistically equivalent to the historical ones and the seasonality in $ET_c$ was compatible to that in forests in the northern hemisphere as reported in Fisher et al. (2011). The maximum C protection capacity at different soil depths at the Coulissenhieb (I) site are available in Guggenberger and Kaiser (2003) and were used to estimate $Q_{max,C}$ in Eq. (4). However, no equivalent data were available for N; hence, a texture-specific protection capacity was estimated using the empirical relationships by Alshameri et al. (2018) for $Q_{max,NH_4^+}$, and Black and Waring (1979) for $Q_{max,NO_3^-}$ and $Q_{max,NO_2^-}$. All protection capacities are expressed in [g-N(p)/kg-soil].

In addition to the Coulissenhieb (I) site, other three case studies were used for our analysis selected among grasslands in tropical, temperate, and semi-arid climates in Australia (Table 2). These case studies are chosen from previously published sites (Ahrens et al., 2015; Tang et al., 2019) and allow performing target comparison of the behavior of ecosystems with *(i)* different vegetation inputs for similar climates (temperate forest *vs* temperate grasslands) and *(ii)* similar vegetation inputs in different climatic regions (temperate grassland *vs* tropical and semi-arid grasslands). All four case studies do not differ in the biogeochemical parameters values. The differences of vegetations and climates are embedded in the model through the imposed boundary conditions. Forests are notoriously characterized by tall-tree vegetation and more profound roots compared to grasslands. This is modeled in BAMS2 through a different root exponential profile. As such, roots in forests are able to uptake a larger amount of nutrients which is traced by means of a fictitious compound labeled $N_{plant}$ (see Online Resources in Table 4). As a consequence, the nutrients released by vegetation litter in forests will be more abundant proportionally to root distribution and amount of nutrient uptaken. By varying the climate, we instead vary the precipitation and its daily distribution (which is in our case studies 0.80 mm day$^{-1}$ for semi-arid, 1.05 mm day$^{-1}$ in temperate and 3.29 mm day$^{-1}$ in tropical grasslands) and, thus, the humidity and saturation of soil. A higher humidity increases the availability and concentration of dissolved compounds in soil which can be readily uptaken by vegetation or by bacteria. As such, the climate variation will be reflected on the nutrients uptaken and released by vegetation (traced by $N_{plant}$). In grassland case studies, the soil profile was extended to 2 meters depth with free drainage bottom boundary condition. The textural fractions and stratigraphy at each site were taken from the SoilGrid database (Hengl et al., 2017) and were used to estimate the hydraulic properties as before. Likewise for the Coulissenhieb (I) site, we generated synthetic precipitations and potential evapotranspiration time sequences using historical precipitations from 1979 to 2017 obtained from the CPC US Unified Precipitation data by NOAA/OAR/ESRL PSD, Boulder Colorado, USA (Xie et al., 2010) and temperature from the Global Historical Climatology Network-daily dataset (Menne et al., 2012) . $ET_c$ for grasslands was obtained using $k_c = 0.8$ after Allen et al. (2005). No data of maximum C and N protection capacities were available for those soils; hence, the maximum C protection capacity was estimated with the empirical relationship in Six et al. (2002) while the maximum protection capacities for inorganic N were estimated as described before for the Coulissenhieb (I) site.

The time horizon considered in our simulations was 380 years. All analyses were carried out using modeling results of the last 100 years, the time during which the SOM profiles approached a near-stationary state.

For simplicity, the Coulissenhieb (I) site will be referred to as TEF while the temperate, semi-arid and tropical grasslands will be referred to as TEG, SAG and TRG, respectively.

## 3 Results and Discussion

### 3.1 Model Benchmarking

Figure 2a shows that BAMS2 captured the observed $C_{(tot)}$ and $C_{(p)}$ profiles relatively well (solid black and red lines, respectively) even if the majority of the biogeochemical parameter were taken from the existing literature rather than being estimated against those data. We compared our model predictions against those by the COMMISSION model (dashed lines in Figure 2; Ahrens et al., 2015), which was specifically calibrated against those data, and we found that the two SOM models were in close agreement relative to the profiles. We note that COMMISSION was run with 100 nodes over 80 cm soil depth and the maximum C protection capacity profile was estimated from data; in contrast, the lower spatial resolution in BRTSim led to greater profile segmentation as compared to COMMISION but $Q_{max,C}$ in BRTSim was assumed to vary over the soil depth as a function of soil composition and it was assigned *a priori* rather than estimated by inverse problem solving (see Section 2.4).

The AER biomass profile predicted by BAMS2 was in good agreement with the few observation points available at Coulissenhieb (I) site (solid black line in Figure 2b) as compared to the overestimation by COMMISSION.

The organic N was generally predicted well except for a slight overestimation in the Bh and Bs horizons as compared to observations (Figure 2c). Possibly, the N profile can be improved by more rigorous estimation of SOM protection parameters and Pgl depolymerization rate, which is characterized by high N content (i.e., low C:N ratio). Relative to organic N, no comparison with COMMISION predictions was possible because it does not account for the N cycle.

Overall, the BAMS2 SOM biogeochemical model integrated into the BRTSim solver was able to reproduce observed SOM and biomass profiles with equivalent accuracy as compared to the COMMISSION SOM model calibrated against the same experimental data. However, the BAMS2 SOM model provides additional capabilities to describe the N biogeochemical cycle and assess nutrient dynamics in vegetated ecosystems.

## 3.2 Results of Sensitivity Analysis

The parameter screening was conducted on the TEF site and on the independent TEG, SAG and TRG grassland sites described in Section 2.4. Because we retained the same BAMS2 biogeochemical parameters employed for TEF, we first verified that BRTSim met the total organic C reported in the SoilGrid database (Hengl et al., 2017).

Computed $\mu$ indices (Eqs.(5)-(6)) for the selected model outputs (Section 2.3) are reported in Figure 3 for TEF (panels a and e), TEG (panels b and f), SAG (panels c and g), and TRG (panels d and h). Results associated with organic C stock (C-stock), organic N stock (N-stock), and inorganic N stock ($N_{inorg}$-stock) are displayed in Figure 3a-d while $CO_2$ and $NH_3$ gas emissions are grouped in Figure 3e-h. The results associated with $N_2O_{(g)}$ and $NO_{(g)}$ are omitted because the $\mu$ indices were not reliable for these target outputs. The latter were less than $10^{-10}$ kg-N day$^{-1}$ m$^{-2}$ when AOB, NOB and DEN were not active for some parametric combinations. This condition was verified for more than 50% of the sampled trajectories. Hence, the proposed formulation of $\mu$ indices (see Eq. 6), which relies on the median statistics, resulted not to be appropriate to investigate $N_2O_{(g)}$ and $NO_{(g)}$ sensitivity.

Figure 3 highlights that the $\mu$ index associated with the AER mortality rate $\delta_{AER}$ showed an high value against all target outputs in the four test sites. This result is consistent with the Monod kinetics in Eq. (1), i.e., slower AER mortality increases the net AER growth rate d[*AER*]/dt and favours AER abundance. Since AER constitutes the most relevant soil microbial population, increases in AER lead to greater $CO_2$ respiration. $\delta_{AER}$ affects also the $N_{inorg}$-stock and the $NH_{3(g)}$ emissions because $NH_4^+$ is a substrate for AER growth and it is the major N pool in $N_{inorg}$-stock in all ecosystems (ranging between 50 to 95 w-N% of aqueous $N_{inorg}$-stock). Since $NH_4^+$ is in equilibrium with $NH_3(g)$, changes in $NH^+$ are then reflected by $NH_3(g)$ emissions highlighting the close link between $CO_2$ and $NH_3$ emissions and AER activity (see later analysis in Section 3.5). The variability of $CO_2$ and $NH_3$ emissions can be therefore considered as a proxy of AER growth as confirmed by earlier results in, e.g., Postma et al. (2007).

We have identified instances where the contribution of AER mortality to the outputs sensitivity was not associated with the largest $\mu$ index among all: (*i*) $\mu$ of root exudates rate $k_{root}$ overcame $\mu$ associated with $\delta_{AER}$ for the N-stock in TEF (Figure 3a); and (*ii*) the nitrogen fixation rate $k_{N-fix}$ in grasslands was associated with the largest $\mu$ for the C and N stocks (Figures 3b,c and d). These instances are discussed in greater details in the following sections, where we highlight the differences that emerge when comparing sensitivity analysis results for different vegetation inputs and climates.

## 3.3 Sensitivity differences across vegetation inputs

Figure 3 shows that the target outputs in the forest (panels a and e) were highly sensitive to the rates $k_{root}$, $k_{leaf}$ and $k_{wood}$ of plant litter and root exudate inputs to soil, but these were not sensitive to the N fixation rate $k_{N-fix}$. The opposite result was found for the three grasslands (panels b-d and f-h). This diverse parameter sensitivity reflects a key difference in factor controlling N availability in soils of different vegetation cover. Our results show that the bacterial and fungal growth in grasslands is very sensitivity to N fixation rather than vegetation inputs, consistently with the literature (e.g., Fornara and Tilman, 2008; Hefting et al., 2005) according to which the supply of N in herbaceous ecosystems (such as grasslands) is typically limited by N atmospheric fixation as the vegetation litter is scarcer and poorer in N content than that in forests while vegetation inputs are the major source of nitrogen in forest soils. As such, parameters associated with vegetation inputs (i.e. $k_{root}$, $k_{leaf}$ and $k_{wood}$) result to be largely influential on SOM dynamics. In addition, note that the pivotal role of N-fixation in providing N to soil is similar in all three grasslands, regardless of the climate. This result is consistent with the work of Hefting et al. (2005) who concluded that climate shifts do not influence the N sources in soil.

The results of our analysis allow a preliminary assessment of the relative importance of different monomers in driving gaseous emissions. On the one hand, AmA and Msa maximum consumption rates by AER ($k_{Resp-AmA}$ and $k_{Resp-Msa}$, respectively) are influential parameters for $CO_2$ and $NH_{3(g)}$ emissions across all test sites, which included different vegetation cover and climates (see Figure 3). On the other hand, $k_{Resp-Phe}$, $k_{Resp-Lip}$, and $k_{Resp-OrA}$ are influential parameters in the forest (Figure 3a) but not in the three grasslands (Figures 3b,c and d). We can support this finding by inspecting the relative abundance of the SOM pools grouped by polymers, and protected/aqueous monomers in the four sites computed by averaging their integral mass in the first 30 cm of soil (pie charts in Figure 4). The total organic C in the aqueous phase (Tot $C_{(aq)}$) was reported as the average C mass fraction [g-C $kg_{soil}^{-1}$] along with its distribution within the monomer pools. Here, dissolved SOM contributed to 25% of the total SOM in the forest while it was always smaller than 5% in the grasslands. The small dissolved SOM pool in grassland was almost entirely made by Msa and AmA, which are the most important substrates for microbial growth. Consequently, change in AER uptake of AmA and Msa has a large impact on the variability of $CO_2$ and $NH_{3(g)}$ emission rates. The partition of dissolved SOM in forest (Figure 4a) highlights that Msa was the most abundant monomer (2.06 g-C $kg_{soil}^{-1}$) followed by Lip (0.554 g-C $kg_{soil}^{-1}$), AmA (0.533 g-C $kg_{soil}^{-1}$) and Phe (0.161 g-C $kg_{soil}^{-1}$). The OrA$_{(aq)}$ in forest ($10^{-2}$ g-C $kg_{soil}^{-1}$) was at least two orders of magnitude greater than in the three grasslands. Increased availability in a monomeric SOM pool corresponded to higher consumption rates and greater $CO_{2(g)}$ respiration as per Eq. (2). Note that a larger amount of Lip, Phe and OrA is the result of the role played by plant litter inputs in forests (as discussed above in this Section) - we recall that these SOM pools constitutes root exudates (OrA and Lip), wood (Phe and Lip) and leaf (Lip) litter, or they are products of lignin depolymerization (Phe) (see reactions R54, R55 and R56 in Online Resource).

## 3.4 Sensitivity differences across climates

The relative and absolute abundance of dissolved SOM in grasslands increasing from the semi-arid (SAG, 1% and Tot $C_{(aq)} = 3.7 \times 10^{-2}$ g-C $kg_{soil}^{-1}$ respectively) towards the wetter tropical grassland (TRG, 5% and Tot $C_{(aq)} = 0.397$ g-C $kg_{soil}^{-1}$, respectively) suggest that climate may act as a driver for SOM accumulation (Figure 4). The available assimilable dissolved SOM impacts on the average $CO_2$ emission rate, which is the lowest in SAG ($7.06 \times 10^{-5}$ kg-C m$^{-2}$ day$^{-1}$ across all simulations) and the highest in TEF (0.0012 kg-C m$^{-2}$ day$^{-1}$).

A non-negligible sensitivity of $NH_{3(g)}$ emission rate to AOB nitrification was identified in TEG and SAG but not in TRG (Figure 3f, g and h, respectively). This suggests that nitrification contributed to model output variability in a different way depending on the climate under equivalent vegetation inputs. We have further investigated this aspect: Figure 5a displays the mean water saturation profile ($S_l$, [-]) in the three grasslands, while Figure 5b shows the corresponding average AOB and AER profiles in [kg-C $m_{soil}^{-3}$]] yielded by the SOM model using benchmark parameter values (see Table 5, Online Resource). We found that the AER profiles were nearly equivalent regardless of the climate, with the greatest AER concentration near the soil surface. Note that AER in the TRG dropped faster as a function of soil depth than in TEG and SAG probably because a high soil water saturation inhibits AER growth as described by the microbial water stress function accounted for in BAMS2 (see further details in Online Resource, Section S2). On the other hand, the AOB mass was not negligible only in SAG and TEG root zone (i.e., greater than $10^{-10}$ kg-C $m_{soil}^{-3}$). Possibly, AOB growth was prevented by the rainfall regime associated with the tropical climates, which maintains a high soil water saturation (Figure 5a). Indeed, intense precipitation in tropical grassland throughout the year caused the rapid flushing of $NH_4^+$ to below the root zone, thus decreasing $NH_4^+$ availability to AOB for nitrification. In contrast, the succession of dry periods in SAG and TEG slowed down $NH_4^+$ leaching and allowed for AOB growth. This explains the sensitivity of $NH_{3(g)}$ emissions to $k_{Nitro-AOB}$ (maximum degradation rate of $NH_4^+$ by AOB) and $K_{M,NH_4^+(AOB)}$ (half-saturation constant in the Michaelis-Menten equation). That is, a greater AOB presence leads to faster $NH_4^+$ consumption and lower $NH_3(g)$ emissions. The absence of AOB in TRG was further explained by the inorganic N pools (Figure 5c, d and e for the SAG, TEG, and TRG sites). Here, $NH_4^+$ was the major component of dissolved $N_{inorg}$-stock in SAG and TEG, followed by $NO_2^-$ and $NO_3^-$. N partitioning in TRG (Figure 5e) was coherent with the absence of AOB (Figure 5b); that is, nearly all $N_{inorg}$-stock consisted of $NH_4^+$ and the NOB activity was inhibited by a lack of $NO_2^-$ production by AOB.

The $NH_{3(g)}$ emissions were also sensitive to AOB mortality, which controls the net AOB growth rate together with $k_{Nitro-AOB}$ as in Eqs. (1)-(2). We also note that a small AOB concentration in the top soil was partially explained by the half-saturation constant of $NH_4^+$ for AOB being greater than the one for AER, thus meaning that AOB were outcompeted by AER when $NH_4^+$ was limiting such as in the top soil at both TEG and SAG sites.

$NH_{3(g)}$ emissions as well as other target outputs were not sensitive to denitrification parameters (Figure 3), but we do not have evidence of how these may be influential on other target variables not investigated in this work.

## 3.5 Sensitivity similarities across climates and vegetation inputs

The $\mu$ indices have identified a non-negligible sensitivity of $NH_{3(g)}$ but not of $CO_2$ emissions (Figure 3). This is a consequence of the $CO_2$ and $NH_3$ solubilities in water (Blanc et al., 2012, see, e.g., Thermoddem database)

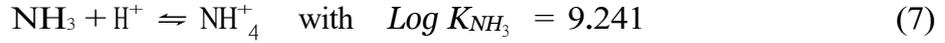
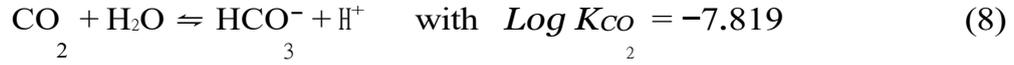

$$NH_3 + H^+ \rightleftharpoons NH_4^+ \quad \text{with} \quad Log\ K_{NH_3} = 9.241 \tag{7}$$

$$CO_2 + H_2O \rightleftharpoons HCO_3^- + H^+ \quad \text{with} \quad Log\ K_{CO_2} = -7.819 \tag{8}$$

where $K_{CO_2}$ and $K_{NH_3}$ are the corresponding equilibrium constants. $Log\ K_{CO_2} < 0$ and $Log\ K_{NH_3} > 0$ suggest that $CO_{2(g)}$ gas ex-solution is favoured while $NH_{3(g)}$ dissolution in water is favoured. Hence, $CO_2$ emissions are limited by biogenic $HCO_3^-$ production while $NH_{3(g)}$ emissions are controlled by $Log\ K_{NH_3}$.

The SOM C-stock was the only target output found sensitive to F mortality ($\delta_F$), the maximum degradation rate of polymers (with the exception of Pgl) and their half-saturation constants (Figure 3). Specifically, the contribution of $\delta_F$ to C-stock variability was comparable to the one of $\delta_{AER}$ because the corresponding $\mu$ were similar (Figure 3). In contrast, $CO_2$ emissions were not influenced by $\delta_F$ (Figure 3). The result returned by $\mu$ is counterintuitive because we would expect that $\delta_F$ should affect the $CO_2$ emissions through aerobic depolymerization reactions. Hence, we investigated the relative $CO_{2(g)}$ emission increments as a function of $\delta_{AER}$ and $\delta_F$. For this purpose, we ran BAMS2 for the TEG site, as a showcase, using the benchmark $\delta_{AER}$ and $\delta_F$ values and we calculated the corresponding $CO_2$ emissions, C-stock, the mass of polymers (Pgl, Lig, HCls and Cls), and the protected and aqueous monomers. Next, we varied $\delta_{AER}$ value up to ± 40% and we computed the same target outputs. This procedure was replicated also for $\delta_F$. Results in Figure 6a and b show that an increasing $\delta_{AER}$ and $\delta_F$ corresponded to smaller $CO_2$ emissions. However, this decrease was steeper for $\delta_{AER}$, suggesting that a small change in $\delta_{AER}$ results in a larger change in $CO_{2(g)}$ emission rate than from $\delta_F$.

Hence, the $\mu$ indices identified the relative non-influential contribution of $\delta_F$ given the much higher contribution of $\delta_{AER}$ to $CO_{2(g)}$ emissions variability.

In all sites, the Lig depolymerization maximum rate ($k_{Depo-Lig}$) and Lig half-saturation constant ($K_{M,Lig}$) had a more significant impact on F net growth as compared to the corresponding parameters associated with HCls and Cls depolymerization (see Figure 3 as a consequence of a Lig pool size greater than other polymers (Table 3). By comparing panels c and d in Figure 6, we found that $\delta_{AER}$ and $\delta_F$ produced a variation in C-stock of similar order of magnitude but with opposite sign, i.e. increasing $\delta_F$ increases the C-stock while increments of $\delta_{AER}$ lead to decreasing of C-stock. We explain this as an indirect feedback on SOM depolymerization resulting by a change in the F:AER ratio driven by AER mortality. In specific, Figure 7c shows that F are more abundant than AER in the topsoil, while the opposite is observed for lower $\delta_{AER}$ values (see Figure 7a and b) where AER are prevalent throughout the entire soil profile. Increasing the $\delta_{AER}$ likely leads to a reduced competition for substrates and $O_2$ between the two microbial functional groups, thus favoring F growth and Lig depolymerization, and leading to the increase of F:AER ratio. The Lignin concentration was higher when bacteria prevailed in the topsoil (Figure 6d). The dynamics regulating F:AER prevalence in soil are still a largely debated topic in literature (Fabian et al., 2017; Strickland and Rousk, 2010; Thiet et al., 2006; Bailey et al., 2002; De Vries et al., 2006), but our work shows that a switch from bacterial to fungal prevalence in the topsoil is essentially controlled by AER mortality. On the one hand, our model predicts that F prevalence leads to small decrements of the C-stock. On the other hand, even if a clear experimental evidence of this effect is needed, different authors (e.g., Strickland and Rousk, 2010, and references therein) have suggested that F prevalence should eventually lead to a C-stock increase because fungal necromass is more recalcitrant than the bacterial one and because F tends to increase the soil protection capacity.

## 3.6 Overall comments

Our findings and the existing literature highlight the need to further explore the following aspects.

Bacterial mortality is commonly modeled in simplistic ways such as by first-order kinetics and, therefore, it is a function of only one parameter (German et al., 2012; Wang et al., 2014). Our results showing the high contribution of bacterial mortality in controlling C cycle suggest that such approach may oversimplify other contributing factors including the microbial biodiversity, aging, competition, symbiosis, environmental conditions, and quality and quantity of available substrates or adaptation to external environmental stressors (see, e.g., Howell et al., 1996; Bressan et al., 2008; Schimel et al., 2007; Van Elsas and Van Overbeek, 1993, and references therein) that determine bacterial mortality. Recent studies have proposed more sophisticated approaches in modeling microbial mortality, such as, by linking thermodynamics and biochemical kinetics theories to describe mortality rate as an explicit function of temperature (Maggi et al., 2018), by including density-dependent microbial turnover (Georgiou et al., 2017), and by accounting for drought tolerance traits (Allison and Goulden, 2017). Still, bacterial mortality remains under-represented in modeling and a matter of large debate. We then recommend that mathematical models entailing dynamical response of microbial populations to environmental stimuli must always be supported by an appropriate and robust uncertainty quantification. Indeed, a deterministic SOM model may not be able to capture the complexity and to account for the unavoidable lack of knowledge associated with dynamics of soil natural phenomena, similarly to all biogeochemical and reactive processes occurring in the subsurface environment (see e.g., Crawford, 1999; Bethke, 2007). Overall, the actual importance of bacterial mortality in driving the shift from fungal to bacterial prevalence (and *vice versa*) is highlighted in our analyses but still needs to be verified through laboratory and/or field experiments. However, our analyses appear a promising tool for speculation on possible parameters controlling SOM dynamics to drive the design of experiments and test potential hypotheses on the feedback between AER mortality, F:AER ratio, and C-stock variability.

Model reduction is typically considered after screening analyses. Our sensitivity analysis has shown important differences across different case studies. These differences pinpoint that a robust sensitivity analysis is a mandatory step to support future model reduction procedures. Indeed, our investigations show that model sensitivity may be remarkably dependent on boundary conditions and the choice of target output variables. Hence, model reduction driven by the sensitivity analysis performed on a specific case study may not be straightforwardly exported to different case studies; rather, our results support the need of a tailor-made model reduction accounting for the specific output of interest, model formulation and system boundary conditions. Hence, sensitivity analysis can play a key role to aid in designing model reduction around the scope of the reduced model.

# 4 Conclusions

We have investigated the sensitivity of the soil organic matter (SOM) dynamics to C-N biogeochemical parameters across different climates and for various vegetation inputs using the mechanistic C-N coupled reaction network model BAMS2. The ability of BAMS2 to reproduce realistic SOM profiles was benchmarked against data sampled in a temperate forest located in Bavaria, Germany. Then, a parametric screening was conducted using BAMS2 on the temperate forest used for benchmarking and three Australian grasslands located in semi-arid, temperate, and tropical climatic regions. From the analyses completed in this work, we concluded that $CO_{2(g)}$ emissions, the SOM stock, and the fungi to bacteria ratio in the top soil are mostly sensitive to the mortality of aerobic bacteria than other biogeochemical parameters accounted for in BAMS2. In particular, the mortality of aerobic bacteria is a crucial driver for the C and N cycle in soil, and it controls the variability of both C and N accumulation and emissions regardless of vegetation inputs and climates. Our work also suggests that bacterial mortality is an important driver of the fungal/bacterial ratio and this needs to be verified and tested through experimental evidences. Major differences depending on the climates and vegetation inputs have also been highlighted by our analyses: (*i*) root exudates and vegetation inputs control the soil nitrogen availability in forests, while this is controlled by atmospheric fixation in grasslands; (*ii*) the higher dissolved SOM content in forest, caused by the abundant vegetation inputs, leads to $CO_{2(g)}$ emissions greater than in grasslands; and (*iii*) the climate and corresponding precipitation intensity might inhibit the activity of specific bacterial functional group such as the nitrifying bacteria (AOB) in tropical climates.

# 5 Acknowledgments

The authors greatly acknowledge the support of Alberto Guadagnini and Giovanni Porta in advising and revising the development of this work. This work is supported by the SREI2020 EnviroSphere, the Mid Career Research, and the Sydney Research Accelerator Fellowship (SOAR) of the University of Sydney. The authors acknowledge the Sydney Informatics Hub of The University of Sydney for providing the Artemis high performance computing resources that have contributed to the results reported within this work. The BRTSim solver package can be downloaded at
https://sites.google.com/site/thebrtsimproject/home
or at the mirror link
https://www.dropbox.com/sh/wrfspx9f1dvuspr/AAD5iA9PsteX3ygAJxQDxAy9a?dl=0.
G. Ceriotti would like to thank the EU and MIUR for funding in the frame of the collaborative international Consortium (WE-NEED) financed under the ERA-NET WaterWorks2014 Co- funded Call. This ERA-NET is an integral part of the 2015 Joint Activities developed by the Water Challenges for a Changing World Joint Programme Initiative (Water JPI).

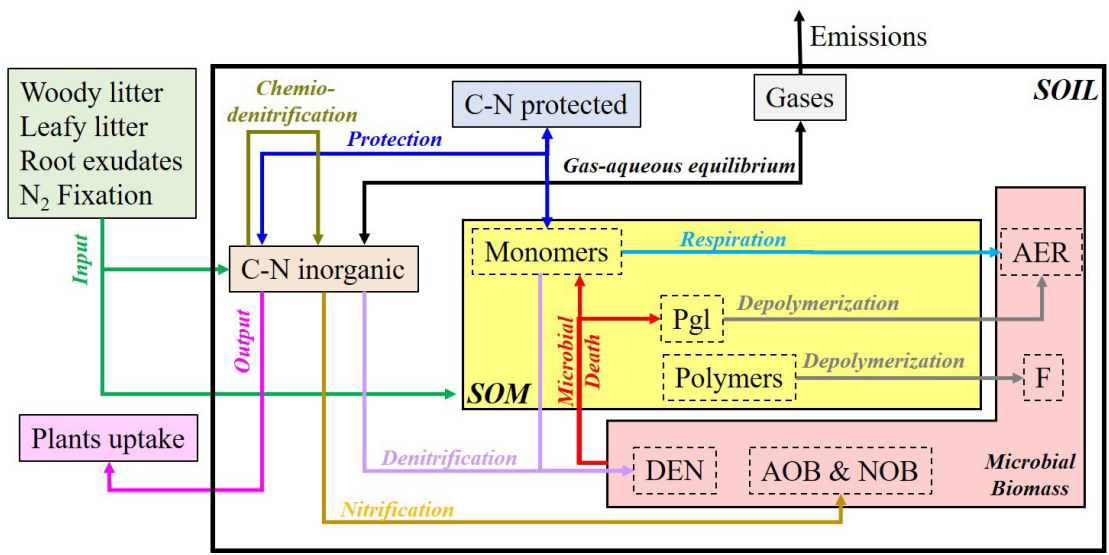

**Figure 1:** Outline of substrate requirements by microbial functional groups for the metabolic processes of nitrification, denitrification, respiration, and depolymerization included in this work. Mass fluxes of necromass, inputs (atmopheric N fixation, roots exudates, woody and leafy litters), outputs (plants uptake and gaseous emissions) and chemo-physical processes (protection, gaseous-aqueous equilibrium and chemo-denitrification) are highlighted by directional arrows. Here, C-N inorganic = inorganic dissolved chemicals containing C and N; C-N protected = inorganic and organic compounds stabilized in soil; SOM = Soil Organic Matter; Pgl = Polypetidoglican; DEN = Denitrifying bacteria; AOB = Ammonia oxidazing bacteria; NOB = Nitrite oxidizing bacteria; AER =Aerobic bacteria; F = fungi.

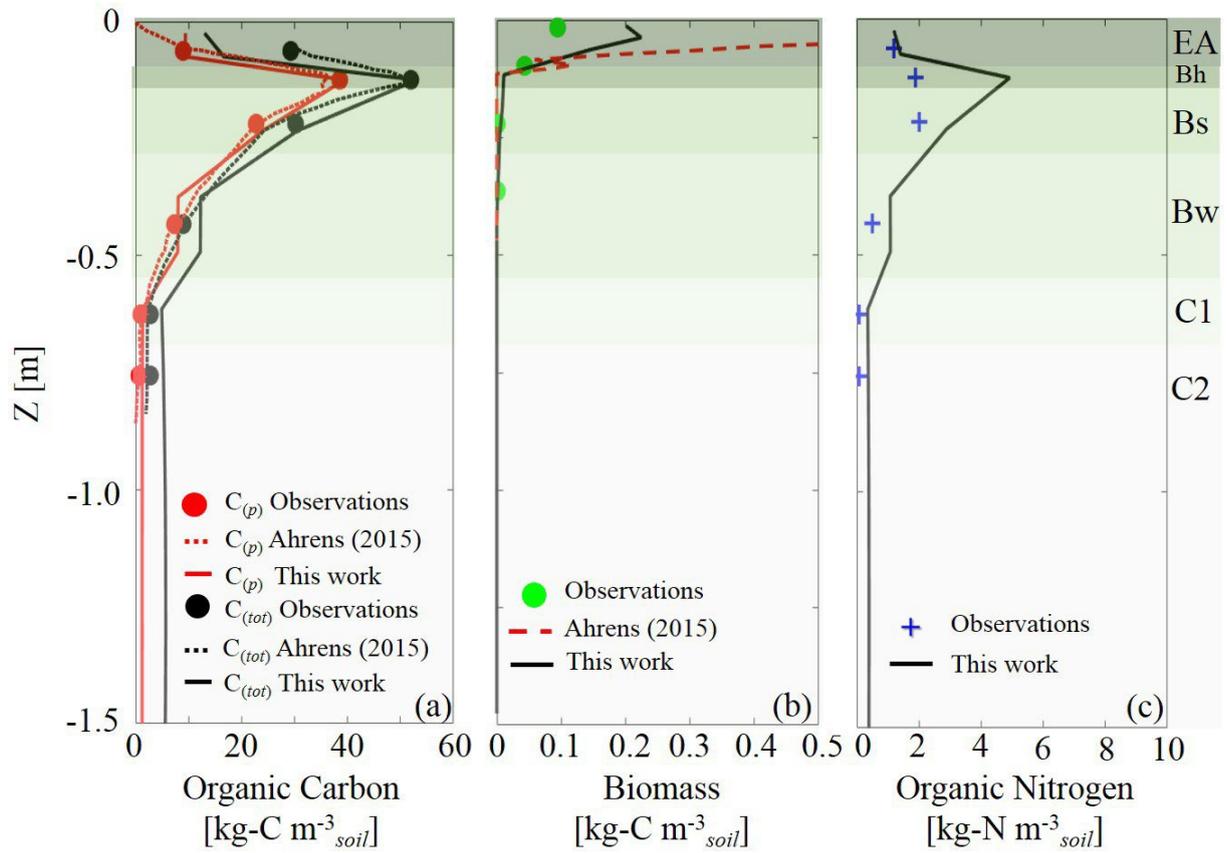

**Figure 2:** Model benchmarking of (a) organic C, (b) microbial biomass, and (c) organic N profiles against corresponding profiles observed at the Coulissenhieb (I) site and profiles predicted by the COMMISSION model (Ahrens et al., 2015) for C and biomass profiles. Shaded areas indicate the horizons stratification (see Table 1). Experimental data are from Staudt and Foken (2007) and Ahrens et al. (2015).

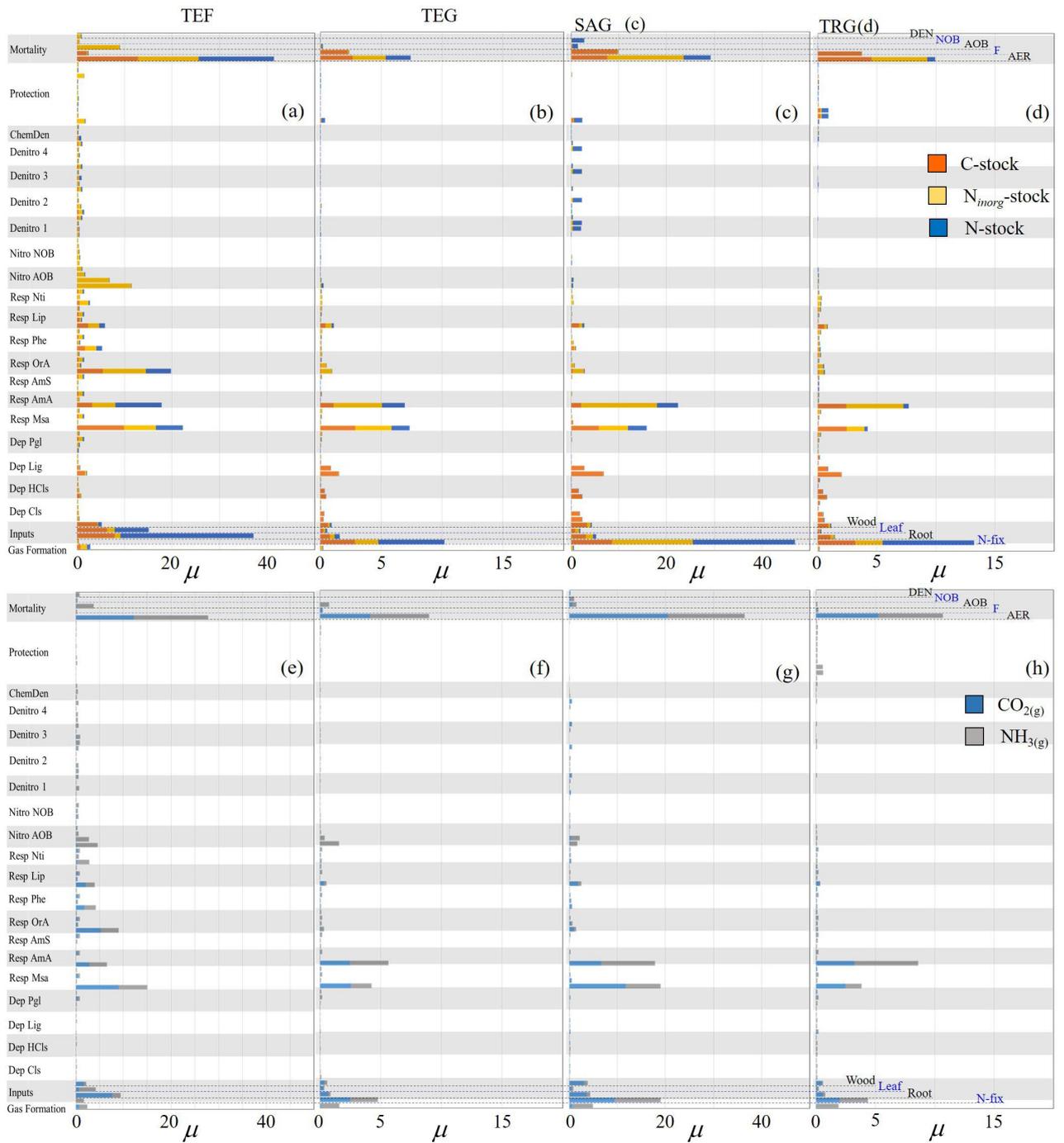

**Figure 3:** Values of the sensitivity (Morris) indices $\mu$ calculated as in Eqs. (5)-(6) for all biogeochemical parameters under scrutiny organized in 22 groups detailed in Table 6, Online Resource, Section S.1 associated with C-stock, $N_{inorg}$-stock, and N-stock for: (a) temperate forest (TEF); (b) temperate grassland TEG; (c) semi-arid grassland (SAG); and (d) temperate grassland (TRG).

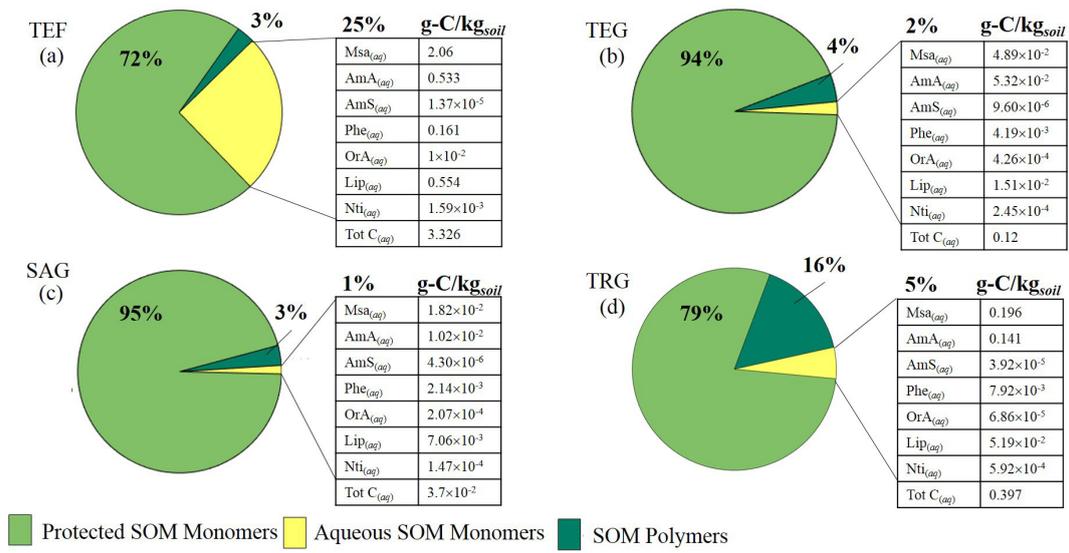

**Figure 4:** Relative abundance of protected and aqueous SOM monomers, and SOM polymers of the C-stock expressed in percent kg-C for: (a) temperate forest (TEF); (b) temperate grassland (TEG); (c) semi-arid grassland (SAG); and (d) temperate grassland (TRG). For each site the The partitioning of aqueous SOM into the different pools (Msa, AmA, AmS, Phe, OrA, Lip, Nti) reports the average g-C/kg$_{soil}$ in each pool in the root zone.

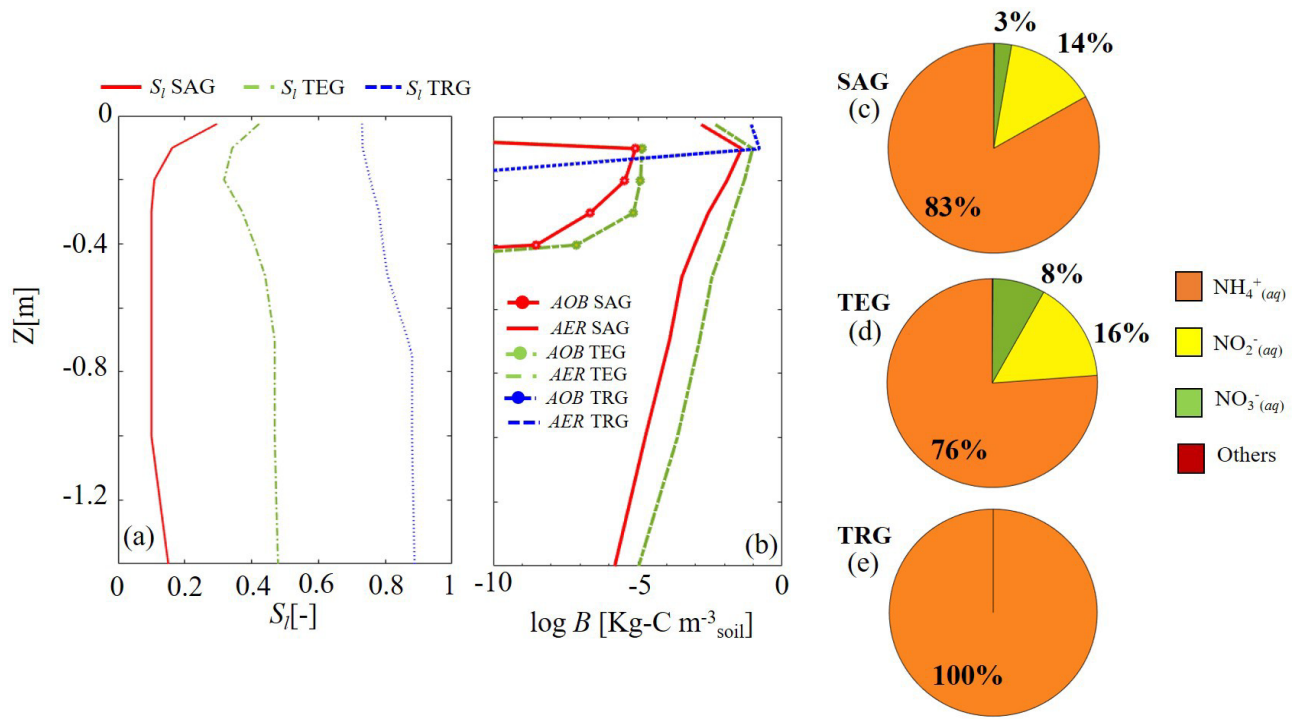

**Figure 5:** Profile of (a) liquid saturation $S_l$, and (b) AOB and AER microbial biomass observed in the semi-arid grassland (SAG), temperate grassland (TEG) and tropical grassland (TRG) calculated by averaging the profiles predicted by BAMS2 over the last 100 years of the simulation period relative to the benchmarked parameter values (see Table 5, Online Resource). Panels (c) to (e) represent the aqueous $N_{inorg}$-stock in inorganic N species for SAG, TEG, and TRG, respectively. Here, B is the density of bacteria in soil expressed in terms of carbon content; AOB = Ammonia oxidizing bacteria; SAG = Semiarid grassland; AER = Aerobic bacteria; TEG = Temperate grassland; TRG = Tropical grassland.

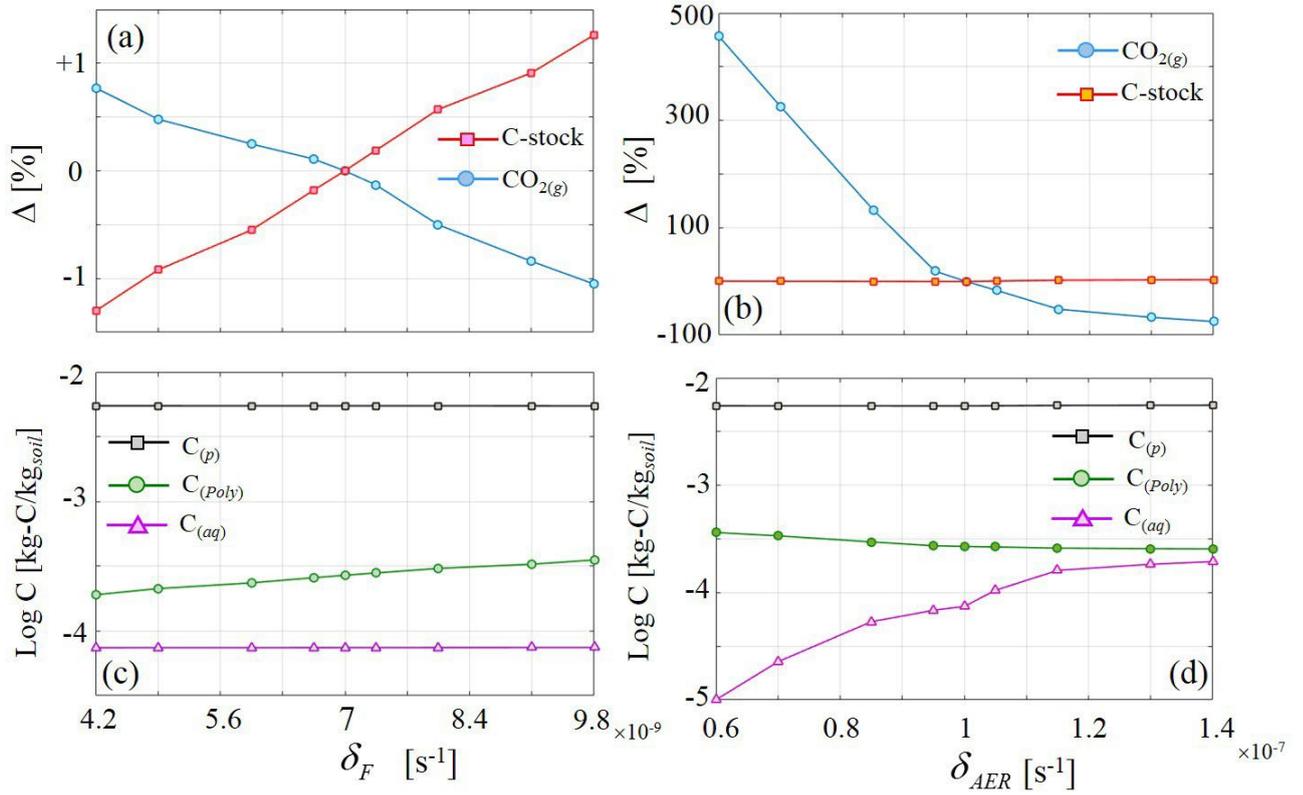

**Figure 6:** Evolution of the relative increment of $CO_2$ emission rate and C-stock with as a function of the mortality rate of (a) aerobic heterotrophs (AER) $\delta_{AER}$ and (b) fungi (F) $\delta_F$ for the temperate grassland (TEG). Evolution of the total amount of carbon present in the soil column partitioned in polymers ($C_{(Poly)}$), aqueous ($C_{(aq)}$) and protected monomers ($C_{(p)}$) as a function of $\delta_{AER}$ (c) and $\delta_F$ (d) for the temperate grassland (TEG).

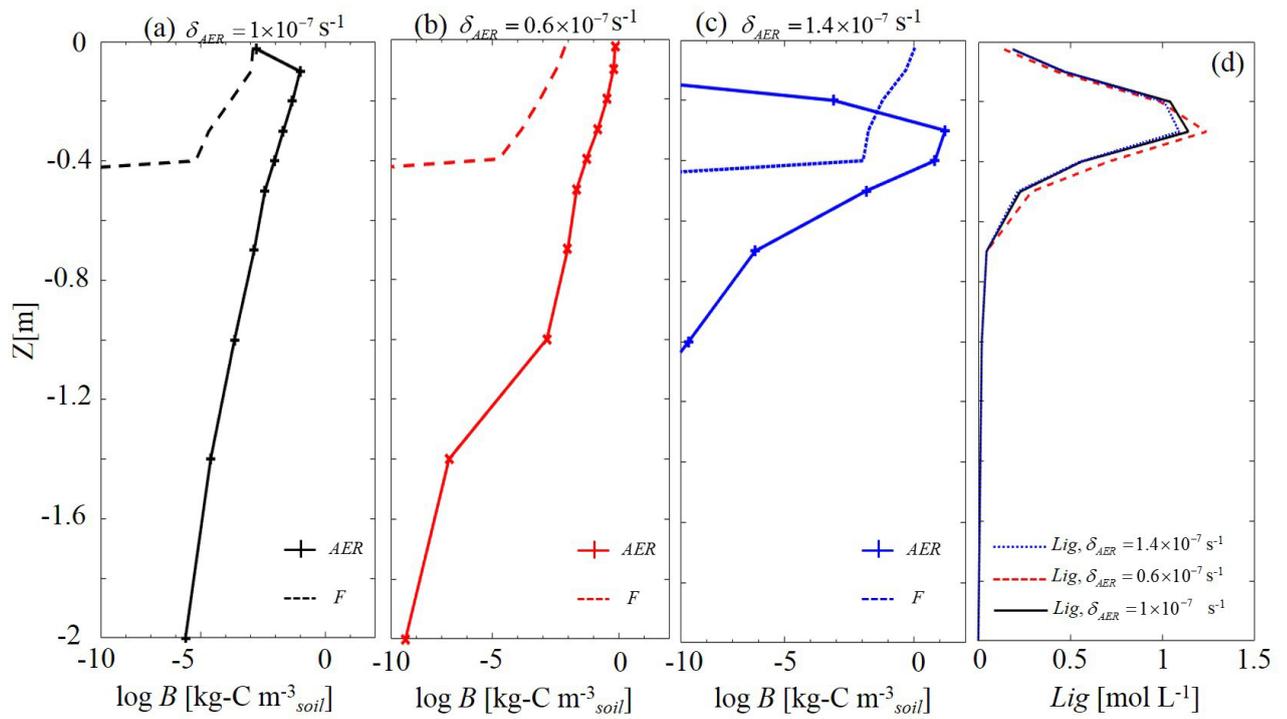

**Figure 7:** Microbial biomass concentration of aerobic heterotrophs (AER) and fungi (F) corresponding to AER mortality rate: (a) $\delta_{AER}$=1x10$^{-7}$; (b) $\delta_{AER}$=0.6x10$^{-7}$; and (c) $\delta_{AER}$=1.4x10$^{-7}$ at the temperate grassland (TEG) site calcuated by averaging the SOM profiles over the last 100 years of simulation. Panel (d) depicts the aqueous concentration profiles of Lignin for $\delta_{AER}$=1x10$^{-7}$ s$^{-1}$, $\delta_{AER}$=0.6x10$^{-7}$ s$^{-1}$ and $\delta_{AER}$=1.4x10$^{-7}$ s$^{-1}$ in the TEG site calculate by averaging the SOM profiles over the last 100 years of the simulation period.

| Layer | Location | Sand [%] | Silt [%] | Clay [%] |
|-------|----------|----------|----------|----------|
| Ea | 0-10 cm | 52 | 38 | 10 |
| Bh | 10-12 cm | 34 | 50 | 16 |
| Bs | 12-30 cm | 45 | 45 | 10 |
| Bw | 30-55 cm | 46 | 43 | 11 |
| C1 | 55-70 cm | 56 | 34 | 10 |
| C2 | >70 cm | 51 | 38 | 11 |

**Table 1:** Stratigraphy and soil composition at Coulissenhieb (I) site (Staudt and Foken, 2007).

| Site ID | Latitude [°] | Longitude [°] | Climate |
|---------|--------------|---------------|-----------|
| TEG | -33.8202 | 135.2551 | Temperate |
| SAG | -29.3240 | 120.2663 | Semi-Arid |
| TRG | -14.3539 | 126.7190 | Tropical |

**Table 2:** Coordinates and climates for the three temperate (TEG), semi-arid (SAG), and tropical (TRG) grasslands in this study. The climate classification is from the modified Köppen climate chart of the Bureau of Meteorology, Australia (Stern and Dahni, 2013).

| Site | Lig w-C[%] | HCls w-C[%] | Cls w-C[%] |
|------|------------|-------------|------------|
| SAG  | 60         | 9           | 31         |
| TEG  | 68         | 9           | 23         |
| TRG  | 76         | 8           | 16         |
| TEF  | 71         | 13          | 16         |

**Table 3:** SOM polymer partitioning into Lignin, Hemi-Cellulose and Cellulose in the root zone of the temperate (TEG), semi-arid (SAG), and tropical (TRG) grasslands, and in the temperate forest (TEF) considered in this study calculated using the benchmark parameter values reported in Table 5 (Online Resources) and averaged on the last 100 years of the simulation period.